\documentclass[aps,prl,showpacs,twocolumn,groupedaddress,floatfix]{revtex4}
\usepackage{amsmath,bm}
\usepackage{mathrsfs}
\usepackage{amsfonts}

\usepackage{graphicx}
\usepackage{xcolor}
\usepackage{ulem}

\makeatletter
\newcommand\colorsout[1]{\bgroup \markoverwith{\textcolor{#1}{\rule[0.5ex]{2pt}{0.4pt}}}\ULon}

\makeatother

\begin{document}

\title{
Correlation-mediated processes for electron-induced switching
between N\'eel states of Fe anti-ferromagnetic chains
}

\author{Jean-Pierre Gauyacq}
\affiliation{Institut des Sciences Mol\'eculaires d'Orsay, ISMO,
 Unit\'e mixte CNRS-Univ
 Paris-Sud,UMR  8214, B\^atiment 351, Univ
 Paris-Sud, 91405 Orsay CEDEX,
France}
\author{Sime\'on Mois\'es Yaro}
\affiliation{Departament de Enginyeria Electronica, Escola Tecnica Superior d'Enginyeria, Universitat Aut\`onoma de Barcelona, Bellaterra, Spain} 
\author{Xavier Cartoix\`a}
\affiliation{Departament de Enginyeria Electronica, Escola Tecnica Superior d'Enginyeria, Universitat Aut\`onoma de Barcelona, Bellaterra, Spain} 
\author{Nicol\'as Lorente}
\affiliation{Centre d'Investigaci\'o en Nanoci\`encia i Nanotecnologia
 (CSIC-ICN), Campus de la UAB, Bellaterra, Spain}
\date{\today}
\begin{abstract}
The controlled switching between two quasi-stable N\'eel states in
adsorbed anti-ferromagnetic Fe chains has recently been achieved by
Loth et al [Science {\bf 335}, 196 (2012)]. In order to rationalize
their data, we evaluate the rate of tunneling electron-induced switching
between the N\'eel states. Good agreement is found with the experiment
permitting us to identify three switching mechanisms: $(i)$ low-bias
direct electron-induced transitions, $(ii)$ intermediate-bias switching
via spin-wave-like excitation, and $(iii)$ high-bias transitions mediated
by domain wall formation. Spin correlations in the anti-ferromagnetic
chains are the switching driving force leading to a marked 
chain-size dependence.
\end{abstract}

\maketitle

The search for nano-scale electronic devices has prompted intense
research in the field of nano-magnetism. The use of spin as
the information conveying entity has steered much excitement due to its
extraordinary properties of information storage, speed and low-energy
consumption~\cite{Wolf2001,Bogani2008,Behin2010}. Miniaturization
is quickly proceeding, reaching very small
domain-wall devices~\cite{Parkin2008}, atomic-size
devices~\cite{Alex2011} and the realm of molecular
devices.~\cite{Bogani2008,Rocha2005,Affronte2008,Mannini2009,Troiani2011} Among
all these possibilities, antiferromagnetically (AFM) coupled devices
have recently received a lot of attention. The AFM characteristics
make these devices very well fitted for quantum computation since they
naturally involved entangled states.~\cite{Meier2003,Cardini2010} Moreover, the
storage in AFM devices is particularly robust due to the lack
of a total magnetic moment. However, this robustness has deterred their use
because changing their magnetic state becomes difficult~\cite{Loth2012}.

Recently,  Loth and co-workers succeeded in controllably switching the
spin states of AFM atomic chains.~\cite{Loth2012} Two quasi-stable
N\'eel states, exhibiting alternating spin directions on the
atoms along the chain, were evidenced in Fe chains adsorbed on a
CuN/Cu(100) {surface}. Loth and co-workers showed that the N\'eel states can be
switched by tunneling electrons injected from a polarized Scanning
Tunneling Microscope (STM) tip into one of the atoms of the chain.
This demonstrated the possibility of storing information on atomic
scale anti-ferromagnet. Theoretical predictions show that writing and
reading spin states entail fundamental problems associated with the
quantum nature of the process.~\cite{Delgado2012} Spin manipulation
by tunneling electrons has been pictured as due to a spin-torque mechanism
where spin angular momentum from the electron is transferred into the atomic spin
system.~\cite{Loth2010,Delgado2010,Novaes2010} However, due
to the lack of magnetic moment in AFM systems, spin manipulation must follow
a different mechanism. Unavoidably, spin manipulations
and excitations are closely related.~\cite{Gauyacq2012}  {Indeed, switching 
between  N\'eel states has been experimentally associated with
overcoming an activation energy.~\cite{Loth2012} 
However, in the case of degenerate  
N\'eel states, resonant  
transitions should be expected.} Hence, the experimental data raise many questions
regarding the possibility of  {resonant}  
switching,  {the efficiency of
activated} 
switching, the nature of the involved excitations
and the physics at play in AFM spin torque. In summary, a complete view
of the switching process is missing.

In this Letter, we reveal the switching mechanisms at play in the
experiment of Ref.~\cite{Loth2012}. The mechanisms turn out to be
rich and closely related with the excitation spectra of the
AFM chain. Their understanding give us a handle on the parameters
controlling the switching process. Here, we show that the correlated
spin nature of AFM quantal chains is at the origin of the 
transition between N\'eel states, and from  this, we deduce the
behavior of the switching rate with respect
to applied bias and  chain size.  Hence, our theory shows that the ability
of switching states are intrinsic to AFM-correlated atomic systems.

One Fe atom on CuN/Cu(100) is characterized by an $S=2$
spin~\cite{Hirjebehedin2007,Lorente2009} with a large magnetic
anisotropy~\cite{Hirjebehedin2007}.  The easy axis lies along the line of
N-atoms of the surface.  Experimental~\cite{Loth2012,Hirjebehedin2006}
and theoretical~\cite{Rudenko2009} evidence show that in a first
approximation, a chain of transition metal atoms on CuN/Cu(100) is 
{an ensemble of weakly interacting atoms such that}
 the Fe chain can be described as a
set of $S=2$ spins with magnetic anisotropy and coupled by 
Heisenberg exchange.  This is partially due to the decoupling
properties of the CuN layer~\cite{Loth2010,Novaes2010,Hirjebehedin2006}
and to the considerable distance between Fe atoms. Hence, the {system
can be described by the magnetic Hamiltonian:} 
\begin{eqnarray}
        H_{0} &=& \sum_{i=1}^{N-1} J \vec{S}_i\cdot \vec{S}_{i+1}
\nonumber \\ &+& \sum_{i=1}^{N} \left ( g \mu_{B} \vec{B} \cdot
\vec{S}_i  + D S_{i,z}^{2} + E (S_{i,x}^{2}-S_{i,y}^{2}) \right ),
\label{hamiltonien} \end{eqnarray} 
where $\vec{S_i}$ is the spin of the atom $i$ ($i = 1, N$) and $S_{i,u}$
its projection on the $u$-axis. $D$ and $E$ are the longitudinal and
transversal anisotropy coefficients ($D < 0$). $\vec{B}$ is a macroscopic
magnetic field applied to the system, along the $z$-axis, similarly to
the experiment~\cite{Loth2012}.

N\'eel states are broken-symmetry solutions of 
Hamiltonian~(\ref{hamiltonien}). Hence, Hamiltonian~(\ref{hamiltonien}) 
 cannot represent
the Fe chains of Ref.~\cite{Loth2012}  
in the absence of an inhomogeneity that breaks the symmetry. 
In the present study,
we introduced a phenomenological term to enforce the N\'eel magnetic structure.
Hence, the magnetic Hamiltonian becomes:
\begin{equation}
H_{Mag} = H_{0} + g \mu_B \vec{B}_{inh} \cdot \vec{S}_{1}.
\label{Mag}
\end{equation}
A small inhomogeneous field,  $B_{inh}$, of $0.1~\; T$ acting on one
of the atoms of the chain (here an end atom) is enough to split the
ground states into two N\'eel-like states (noted 1 and 2) that contain
contributions from many spin configurations. This added perturbation
is indeed small since the two N\'eel states are only $\sim 50 \mu$eV
away from each other. The inhomogeneous term can be thought to be
representative of various effects: small inhomogeneities of the surface,
an inhomogeneity of the external B field,  {dephasing effects} or a field induced by the
polarized tip of the STM~\cite{LothandHeinrich}. Following Loth et \textit{
al.}~\cite{Loth2012}, the parameters in Hamiltonian~(\ref{hamiltonien})
were partially determined by fitting the experimental Fe$_2$ excitation
energies. Hence, similarly to the fitting of the Ising-model parameters
in Ref.~\cite{Loth2012}, we set $E = 0$ and obtained $J = 1.6$
meV and $D = -1.34$ meV~\cite{Note}.  Here we consider even-numbered
chains and the Hamiltonian (\ref{Mag}) was diagonalized in a basis
formed by direct products of local spin states (spin configurations):
$\Pi_k |M (k)\rangle$, where $|M (k)\rangle$ is an eigenstate of the
projection on the z-axis of the local spin at atom k, $S_{k,z}$. For the
longer chains, diagonalization methods specific to sparse matrices had
to be used~\cite{Xavi}.

\begin{figure}
\includegraphics[width=0.45\textwidth]{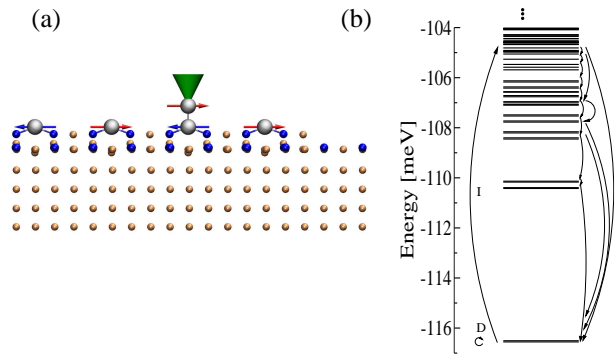}
\caption{
$(a)$ Setup used in the experiment of Ref.~\cite{Loth2012}.
A few atoms of an Fe (large gray balls) monoatomic chain on CuN/Cu(100) are shown
under an STM tip that injects spin-polarized electrons. The Fe-N
bonds are also depicted, the presented arrangement
corresponds to the atomic coordinates of a DFT calculation.
The chain spins alternate along the chain axis to
form a N\'eel state.
$(b)$ Efficient excitation and de-excitation schemes leading
to the switching between N\'eel-states. $D$ refers to the direct
process between the quasi-degenerate N\'eel-states
 and $I$ refers to the indirect one that involves excitation
and de-excitation mechanisms. The up arrow on the left shows the
excitation by a tunneling electron whereas the down arrows on the right
show the relaxation induced by substrate electrons.
All possible de-excitation cascades are included, signaled here by
a few de-excitation paths. The depicted spectrum corresponds
to a chain of 10 Fe atoms.
 \label{figure1}}
\end{figure}

We describe the magnetic excitation during tunneling using the strong
coupling approach of Ref.~\cite{Lorente2009} in which the electron
strongly interacts with the atom it tunnels through.  Let us consider
the system schematized in Fig.~\ref{figure1} $(a)$: the STM tip is
standing above the first atom of a chain in one of the two N\'eel states
and the injected electron is tunneling through it.  Within the sudden
approximation the electron transmission amplitude operator is equal
to:~\cite{Gauyacq2010} 
\begin{equation}
T_{Tip \rightarrow Sub} = \sum_{M_T} |S_T =  5/2, M_T \rangle T^{S_T}_{Tip \rightarrow Sub} \langle S_T =  5/2, M_T |,
\label{T-matrix}
\end{equation}
where $S_T$ is the spin of the compound system formed by the
tunneling electron and the corresponding Fe atom. 
$T^{S_T}_{Tip \rightarrow Sub} $ is the electron transmission
amplitude in the $S_T$ symmetry from tip to substrate.  Here we use
$S_T=\frac{5}{2}$  revealed {as the dominant tunneling channel} by density functional theory
studies in Ref.~\cite{Lorente2009} for an isolated Fe atom on
CuN/Cu(100). The scattering amplitude between the chain magnetic states
is then obtained as the matrix element of the amplitude (\ref{T-matrix})
between initial, $|i\rangle = |\sigma_i\rangle |\phi_i\rangle$  and
final states, $|f\rangle = |\sigma_f\rangle |\phi_f\rangle$, written
as direct products of  the scattering electron state of spin $\sigma$
and  the chain state $|\phi\rangle$, eigenstate of Hamiltonian~\ref{Mag}.
The probability, $P(i \rightarrow f)$, for a transition from the initial
state $|i\rangle$ to the final state $|f\rangle$ associated to the
tunneling of an electron from tip to substrate is proportional
to $|T^{S_T}_{Tip \rightarrow Sub}|^2$ and to $ |\sum_{M_T} \langle
f | S_T = 5/2, M_T \rangle \langle S_T = 5/2, M_T|i \rangle |^2$.
The $|T^{S_T}_{Tip \rightarrow Sub}|^2$ factor corresponds to a global
tunneling probability, whereas the second factor yields the relative
importance of the inelastic channels in the tunneling process. The
(i$\rightarrow$f) probability is very large if there is a strong overlap
between the intermediate state of $S_T$ tunneling symmetry and both
the initial and final states.

\begin{figure}
\includegraphics[width=0.35\textwidth, angle=270]{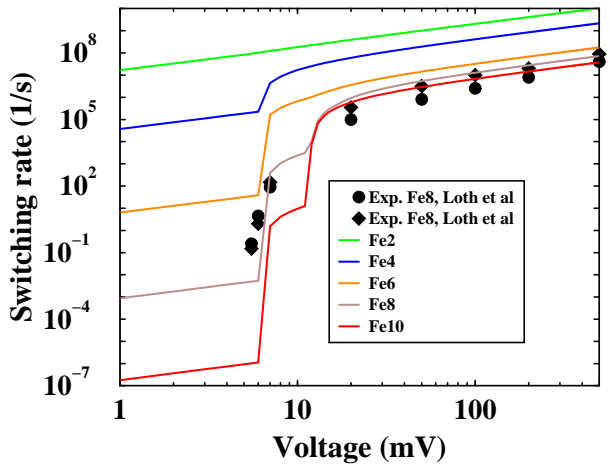}
\caption{
 {Tunneling electron-induced } switching rate between N\'eel states in $s^{-1}$ as a function of the
applied bias in $mV$ for Fe chains of different length. The higher rates
correspond to the Fe dimer, and the rate decreases as the chain length increases.
For comparison the experimental results~\cite{Loth2012} are plotted for transitions from
low-current to high (dots) and high to low (diamonds).
\label{figure2}}
\end{figure}

The electron can only induce a change of one unit in the spin projection
of the atom it tunnels through. As a consequence, in a zero-order view of
the N\'eel states, described as chains of atoms with  alternating spins,
electron-induced switching between pure N\'eel states does not exist.
However, AFM chains described with Heisenberg couplings are strongly
correlated~{\cite{Autre, Gauyacq2011},} i.e.  the two N\'eel-like
states, eigenstates of Hamiltonian~(\ref{Mag}), contain small components
over a very large number of different configurations of atomic spins.
Direct, {quasi-resonant} electron-induced transitions between N\'eel-like
states are then possible. These quantal transitions that do not involve
spin-flip of the tunneling electron are mediated by correlation. However,
since they involve the small components in the state expansion over spin
configurations, their probability is weak. Furthermore, if the length of
the chain is increased, the direct 1$\rightarrow$2 transition probability
decreases rapidly. This is clearly seen in Fig.~\ref{figure2} which
shows the rate for the 1$\rightarrow$2 transition induced by tunneling
electrons as a function of the STM bias. At low bias (below 6 mV),
only direct 1$\rightarrow$2 transitions are possible and lead to a very
small transition probability for long chains that very rapidly decreases
as the chain length increases.  Typically, the direct 1$\rightarrow$2
excitation probability per electron is equal to $0.9 \times 10^{-7}$
($1.5\times 10^{-11}$) for Fe$_6$ (Fe$_8$).

Above 6 mV, the transition rate increases drastically
due to indirect transitions: the system initially in state 1 is excited
by collision with the tunneling electron into a higher-lying state
$i$, and state $i$ later decays into state 2 with a finite probability
(Fig.~\ref{figure1} $(b)$). 
The decay process takes place via electron-hole  {pair} excitation of the 
substrate and it is the inverse of the excitation process discussed
above.~\cite{Novaes2010}
The above strong-coupling approach successfully accounts for excited
spin state lifetimes.~\cite{Novaes2010,Gauyacq2012b} Hence, we
can complete the full excitation/de-excitation dynamics, by considering now
the de-excitation probability $P^{Dec}_{i \rightarrow f}$:
\begin{equation} 
P^{Dec}_{i \rightarrow f} =
\Delta E_{ij} \frac{P(i \rightarrow f)}{\sum_{l < i} P (i \rightarrow l)
\Delta E_{il}}, 
\label{Decay} 
\end{equation}  
where $P(i \rightarrow f)$  has been defined earlier
and $\Delta E_{ij}=E_i - E_f$ is the decay energy. Since there is
no long-lived excited state in the system (except state 2), an excitation
of state $i$, followed by successive decays according to (\ref{Decay}), eventually
results in the population of the two N\'eel states, 
$N_i (1)$ and $N_i (2)$.
 The 1$\rightarrow$2 transition rate for a bias $V$ is then given by:
\begin{eqnarray}
Rate(1 \rightarrow 2) &=& C  [ (eV-\Delta E_{2 1}) P(1 \rightarrow 2) 
\nonumber \\
& + &
\sum_{i>2} (eV-\Delta E_{i 1}) P(1 \rightarrow i) N_i (2)  ].
\label{Rate}
\end{eqnarray}  
The $C$ coefficient is the normalization constant for the total flux
of electrons in a given experiment~\cite{Note2} which is bias independent
for small bias.
The first term in Eq.~(\ref{Rate}) corresponds to the direct transitions already
discussed and the sum over i to the indirect transitions. 

Figure~\ref{figure2} shows the computed rates for Fe$_n$ chains ($n=2-10$)
 compared with the experimental results for Fe$_8$.~\cite{Loth2012}
The theoretical transition rate decreases with chain length and
from there one can conclude that direct transitions are practically
impossible for very long chains whereas the indirect process should be
accessible in a broad range of lengths.  Our results on Fig.~\ref{figure2}
compare very well with the experimental data for Fe$_8$, also shown as
dots and diamonds.  Dots in Fig.~\ref{figure2} show the experimental
transitions rates from low current to high current, and diamonds
from high to low currents.  This corresponds to  $1 \rightarrow2$
transitions and $2\rightarrow 1$ respectively. The experimental
results in Fig.~\ref{figure2} were obtained with a polarized tip,
whereas the theoretical results were obtained with a non-polarized
tip.  Polarizing the tip influences the transition rates, due to the
intermediate S$_{T}$=5/2 symmetry involved in the tunneling. Inelastic
transitions are favored  when the tip and the atom under the tip have
opposite spin directions; the  ratio between the theoretical switching
 rates is typically around 3 for a tip polarization equal to 0.3 which
explains the asymmetry between dots and diamonds in Fig.~\ref{figure2}.

The theoretical switching rates show an abrupt change at $\sim 12$ meV.
This change is more clearly seen in Fe$_8$ and Fe$_{10}$ chains.
By studying the spectra of excitations for these chains, we can
clearly separate two distinct regions in the rate due to two
sets of excitations different in nature.
The first region, between 6 and 12 meV, corresponds to
transitions where $M$ changes in one of the end atoms of the chain; these are very efficient
in shorter chains such as Fe$_6$. This type of excitation is a quantized
spinwave of the finite chain~\cite{Gauyacq2011}. The second region,
beyond 12 meV, corresponds to transitions that mix several configurations
with two opposite anti-ferromagnetic domains, they lead then to
domain-wall formation.  As the chain length increases, domain-wall
mediated transitions become more important than spinwave-mediated
transitions, explaining the clear upturn beyond 12 meV for
Fe$_8$ and Fe$_{10}$. 
 {There is also a dependence of the transition rate on the position of
 the tip along the chain. Indeed, this position effect mixes with the
 polarization effect due to the alternating spin directions along the
 chain in the N\'eel state as well as with the effect of the position of
 the inhomogeneous term (2) along the chain. Besides the interplay with
 the polarization effects, end atoms are more efficient for moderate
 bias and central atoms for high voltages. }

\begin{figure}
\includegraphics[width=0.35\textwidth, angle=270]{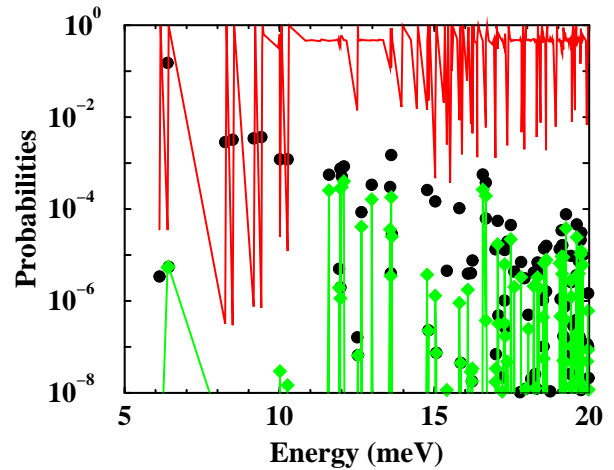}
\caption{
Probability as a function of the $i$-level energy 
for the excitation from
the N\'eel state $1$ to level $i$: $P(1\rightarrow i)$ (black circles); branching
ratio toward state 2 once level $i$ is populated: $N_i(2)$ (red line); 
and total indirect switching probability between N\'eel state $1$ and $2$
via an intermediate $i$-level:
 $P(1\rightarrow i) N_i (2)$ (green diamonds).  Many states $i$ are excited
by an electron, with a probability
roughly exponentially decreasing with the excitation energy, but the
final switching probability is strongly modulated by the de-excitation
probability (branching ratio).
\label{figure3}}
\end{figure}

The transition path $1 \rightarrow i \rightarrow 2$ for Fe$_8$ is analyzed
in Fig.~\ref{figure3}. The different contributions to the switching
rate, Eq.~(\ref{Rate}), are presented as function of the excitation
energy: the primary excitation probability $P (1 \rightarrow i)$ (black
circles), the branching ratio toward state 2 ( $N_i (2)$, red line)
and the total indirect excitation probability  $P (1 \rightarrow i) N_i
(2)$, (green diamonds).  Many states $i$ are excited by a tunneling
electron, though many of them with a small probability that roughly
exponentially decreases with the excitation energy~\cite{Gauyacq2011}.
The largest excitation probability corresponds to states $i$ where the
spin of the atom under STM tip changes by $\Delta M\approx \pm 1$.  All the
other states are excited via \textit{correlation}, i.e. the fact that
states 1 and $i$ are not associated to a single configuration of local
spins~\cite{Gauyacq2011} but to a mixing of a large number of them from
a configuration interactions point of view. As for the decay of the
excited states, $i$, the lower-energy states decay preferentially toward
N\'eel state 1 or 2 depending on which state they are configurationally
closer. For high-lying states, correlation becomes stronger and
the excited states decay roughly equally to the two N\'eel states. 
{Domain walls in particular decay roughly equally to both N\'eel states.
As a conclusion from Fig.~\ref{figure3}, the global transition
$1 \rightarrow i \rightarrow 2$ requires that both excitation and
de-excitation are sizeable, i.e. requires balancing the  $1 \rightarrow i$
and $i \rightarrow 2$ probabilities.  }

{The relative weight of the various switching processes vary with 
chain length and this can be rationalized  considering that the distance
between the two N\'eel states in terms of spin-configuration changes is
increasing with chain length. For the direct quasi-resonant process
at small bias, one needs a strong configuration mixing between the two
N\'eel states and this quickly decreases with  chain length. The
process around 6-8 meV is associated to $\Delta M \approx \pm 1$ transitions
in the atom just under the tip; the corresponding excited state is
still very close to the initial N\'eel state and therefore the decay
to the other N\'eel state is difficult and rapidly decreases with 
chain length increase. The process around 12-13 meV is associated to
domain wall formation and behaves differently. First, several states
contribute and their number increases with chain length; second,
their decay equally populates the two N\'eel states. So, even, if their
excitation probability from the initial N\'eel state is not very high,
they succeed in dominating the indirect process for long chains.  }

In summary, our calculations show three different regimes in the
tunneling electron induced switching of the AFM chains
in Ref.~\cite{Loth2012}.  The low-bias region corresponds to  
{quasi-resonant} direct
transitions between the two N\'eel states.  
The intermediate-bias region
is characterized by the threshold of chain excitations. Beyond this
threshold, {tunneling} electrons induce an indirect process mediated by
spin wave excitations.  A second threshold defines the high-bias
region where domain-wall excitations 
{dominate the switching process}. 
 {Correlation, i.e. mixing of spin configurations in the chain, is
 the driving force of the three N\'eel switching processes. However,
 correlation acts differently in the three, resulting in different
 dependences on the chain length of the three process and leading to
 the dominance of the process involving domain wall formation for long
 chains.}

The mechanisms described in this Letter are very different from the
more usual local spin-flip mechanisms at play in, e.g., magnon
excitation in ferromagnetic chains. Instead,
the present mechanisms should be very general
and operational in many systems with strong correlations, such as
frustrated systems.


\begin{thebibliography}{99}
\bibitem{Note}
Note that the $J$ and $D$ parameters are very close to the Ising representation in Ref.~\cite{Loth2012}
as well as to the fitting of $D$ for a monomer.
\end{thebibliography}
\end{document}